\documentclass[%
 reprint,
showkeys,
 amsmath,amssymb,
 longbibliography,
aps,
prx
]{revtex4-1}

\usepackage{graphicx}
\usepackage{dcolumn}
\usepackage{bm}

\newcommand{\bJ}{\textbf{J}}

\newcommand{\bmcI}{\mcI}

\newcommand{\Tc}{T_{\rm c}}
\newcommand{\Topt}{T_{\rm opt}}

\newcommand{\mcI}{\mathcal I}
\newcommand{\mcL}{\mathcal L}
\def\l{\left}
\def\r{\right}
\def\f{\frac}

\begin{document}


\title{Physically optimizing inference}

\author{Audrey Huang}
\affiliation{Department of Computer Science, California Institute of Technology, Pasadena, California, 91125, USA}
\author{Benjamin Sheldan}
\author{David A.~Sivak}
 \email{dsivak@sfu.ca}
\affiliation{Department of Physics, Simon Fraser University, Burnaby, BC, V5A1S6, Canada}
\author{Matt Thomson}
 \email{mthomson@caltech.edu}
\affiliation{Division of Biology and Biological Engineering, California Institute of Technology, Pasadena, CA, 91125, USA}



\date{\today}

\begin{abstract}
Data is scaling exponentially in fields ranging from genomics to neuroscience to economics. A central question is: can modern machine learning methods be applied to construct predictive models of natural systems like cells and brains based on large data sets? 
In this paper, we examine
how inference is impacted when training data is generated by the statistical behavior of a physical system, 
and hence outside direct control by the experimentalist. 
We develop an information-theoretic analysis for the canonical problem of spin-network inference. 
Our analysis reveals the essential role that the physical properties of the spin network and its environment play in determining the difficulty of the underlying machine learning 
problem. Specifically, 
stochastic fluctuations drive a system to explore a range of configurations providing `raw' information for a learning algorithm to construct an accurate model; yet they also blur energetic differences between network states and thereby degrade information. 
 This competition leads spin networks to generically have an intrinsic optimal temperature at which stochastic spin fluctuations provide maximal information for discriminating among competing models, maximizing inference efficiency. 
We demonstrate a simple active learning protocol that optimizes network temperature to boost inference efficiency and dramatically increases the efficiency of inference on a neural circuit reconstruction task. Our results reveal a fundamental link between physics and information and show how the physical environment can be tuned to optimize the efficiency of machine learning. 
\end{abstract}


\keywords{learning, inference, Fisher information, inverse Ising problem, spin network, control, prediction}
\maketitle


The emergence of `Big Data' is a central theme in contemporary science~\cite{Lichtman:2014fu,Blumenstock:2015hq,ScalingRNASeq}. An important challenge is utilizing large data sets to model and understand complex interacting systems found in fields such as biology and economics~\cite{Shiffrin:2016iz,OnaQuestforPrinc:2016gw}. Recently, machine learning has demonstrated the ability to extract and model patterns from large data sets across domains as disparate as object recognition, speech interpretation, and game playing~\cite{Silver:2016hl,Hinton:2006kc,Schmidt:2009dta,George:2017jr,Bartok:2017hz}. Thus, modern machine learning methods provide an attractive route for automating the construction of scientific models from data.   

In science and engineering, the key objective of modeling is the prediction of system behavior in new conditions. 
For example, an important goal in modern biology is to forward-engineer biological networks in order to achieve new function. Current specific goals include the rewiring of metabolic pathways to synthesize industrial chemicals and cellular signaling pathways to treat diseases like cancer~\cite{Bailey1668}.
To drive networks to states not observed in their natural range of operation, we require network models that can predict the response of a complex biological network to novel perturbations, engineering interventions, or environmental conditions.  

In machine learning, the statistics of training data control the accuracy, generalizability, and efficiency of learning~\cite{vapnik,Valiant:1984kt,Poggio:gt}. A challenge in scientific model construction is that the statistical properties of the training data are determined by the physical behavior of the system under investigation: training data can only be indirectly controlled by human intervention or experimentation. A fundamental research question, then, is to understand how a system's natural statistics affect the learning of predictive models.
In this paper, we ask how the physical behavior of a network determines the feasibility of learning models from observations: when is the natural behavior of a system sufficient for inference? 

We develop an information-theoretic analysis of a classic physical learning problem, the inference of a spin network from observations. Spin networks have been applied to model a range of disordered systems with non-uniform interactions between elements, including neural networks, bird flocks, and economic systems~\cite{Schneidman:2006he,Hinton:2006kc,Hopfield:1982vm,Katz:2011fb,Cocco:2009hw,Tkacik:2009}.  

Our analysis shows that the physical environment (specifically temperature) of a spin network determines when a unique model can be constructed from a finite set of observations. 
We find generically that learning is optimally efficient at an intermediate temperature (and hence intermediate fluctuations):
at low and high temperatures, system fluctuations don’t encode sufficient information to distinguish between many possible competing models; at intermediate temperature the system emits maximal information, making identification of predictive models maximally efficient. 
These insights are borne out in the geometry of the parameter manifold (specifically its curvature), as quantified by the Fisher information.
Through active learning, an observer can optimize the scale of a system's fluctuations, thereby bringing the system to its optimal inference temperature and maximizing the efficiency of inference. 
Thus, our results elucidate a fundamental connection between physics and machine learning by revealing that the physical environment of a network constrains the amount of information that is available for inference.


\section*{\label{sec:SpinNetworkInference}Spin-network inference}
\begin{figure}
  \centering
    \includegraphics[width=0.5\textwidth]{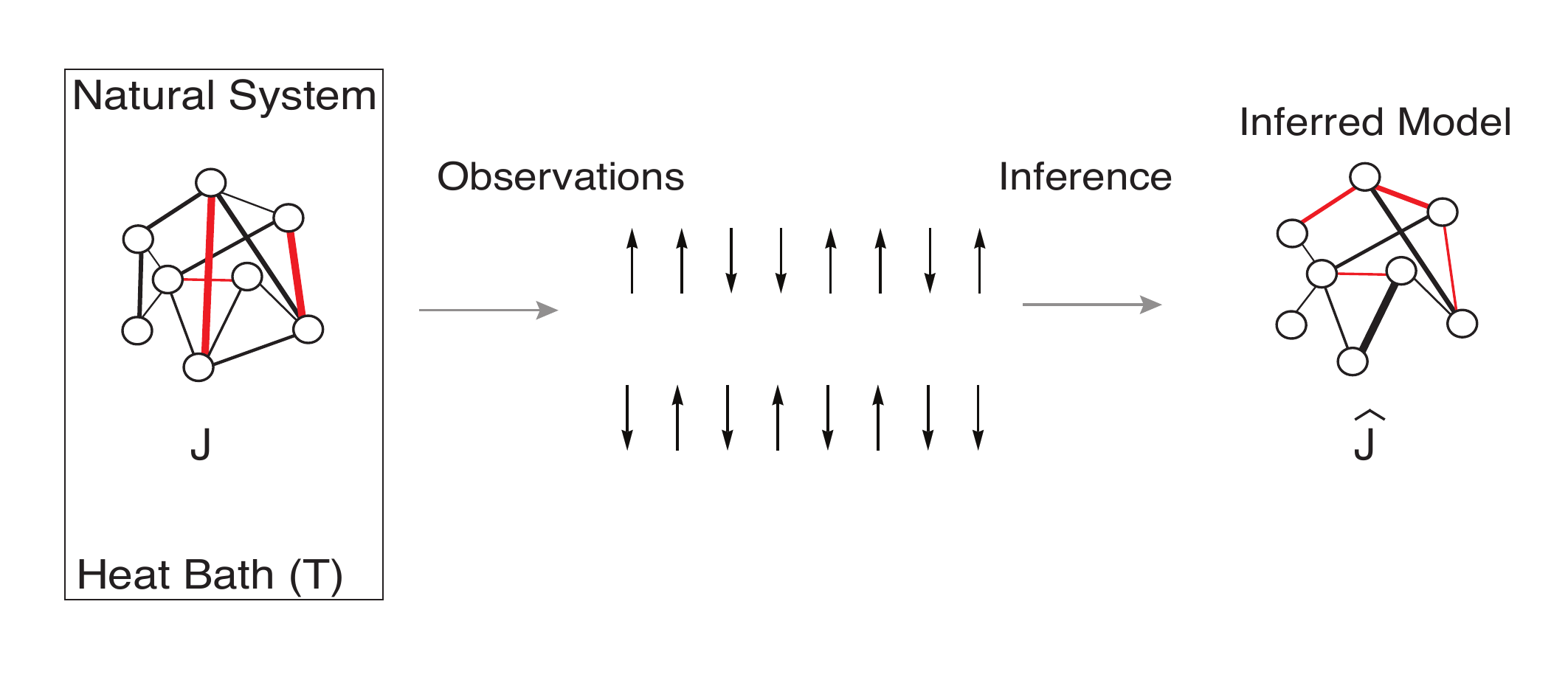}
     \caption{The spin-network inference problem.  A spin network is coupled to a heat bath at temperature $T$. Negative couplings are shown as red links and positive couplings as black links. An observer infers a predictive model of the network based on observations of network configurations drawn from the equilibrium distribution.}
\label{fig:problemSchematic}
\end{figure}

We consider the problem of inferring a predictive model of a spin network from network state observations (Fig.~\ref{fig:problemSchematic}). 
We specifically ask how the physical properties of the network (its physical environment) impact the efficiency of inference.

A spin network consists of $m$ binary elements or `spins.'  A network microstate is specified by each spin $i$ being either up ($\sigma_i=1$) or down ($\sigma_i = -1$).  
(As an alternative to their original physical interpretation, spins can represent Boolean variables that are `true' or `false,' neurons that are `firing' or `silent,' genes that are `on' or `off,' and so on.) Spins $i$ and $j$ interact in a pair-wise fashion parameterized by a coupling constant $J_{i,j}\in \mathbb{R}$. Interactions are symmetric, $J_{i,j} = J_{j,i}$, with no self-interactions, $J_{i,i} = 0$. 
These symmetry and diagonal constraints mean that the matrix $\bJ$ has $(m^2-m)/2$ unknowns.

A given spin network can exist in one of $2^m$ different microstates or configurations, $s = (\sigma_1, \sigma_2, \ldots, \sigma_m)$. Each network configuration has an associated energy due to spin-spin interactions: 
\begin{align}
E(s|\bJ) = -\sum_{ij} J_{i,j} \sigma_{i} \sigma_{j} - \sum_{i} h_{i} \sigma_{i} \ .
\label{energyfun}
\end{align}
A magnetic field $h_i$ biases spin $i$ to either point up or down.  For simplicity, in our analysis we set $\textbf{h}=0$, but the extension to non-zero fields is conceptually straightforward.

The network exchanges energy with a heat bath at temperature $T$. At equilibrium, the network fluctuates between configurations, and the probability of observing a specific spin configuration $s_k$ is determined by the Boltzmann distribution,
\begin{align}
P(s_k|\bJ) = \frac{\exp[-E(s_k|\bJ)/T]}{Z} \ ,
\label{eq:boltz}
\end{align}
for network partition function $Z \equiv \sum_k \exp[-E(s_k|\bJ)/T]$ (throughout we set $k_{\rm B} = 1$). (In more general situations---\emph{e.g.}, a neural network---$T$ would quantify the effective temperature, corresponding to the magnitude of stochastic fluctuations in network state.)  
We observe a given network at equilibrium, sampling $N$ independent identically distributed observations $\{s_k\}$ of network state from the equilibrium distribution. 

Our goal is to learn a predictive model, a model that predicts the distribution $P(s|\bJ)$ of network states under new environmental conditions, for example, at a new temperature $T'$, or in the presence of a magnetic field $\bm{h}$. 
For spin networks, learning a predictive model amounts to learning the spin-coupling matrix $\bJ$ with high accuracy, since the Boltzmann distribution (\ref{eq:boltz}) with given $\bJ$ produces predictions of $P(s|\bJ)$ for arbitrary $T$ or $\textbf{h}$. Therefore, the question becomes how system and environment physical properties (here, temperature) affect the accuracy of the $\bJ$ estimated from sampling network states, and when we can learn $\bJ$ with high accuracy. 

We utilize likelihood-maximization~\cite{Tkacik:2009,cover06,Schneidman:2006he} to estimate the coupling matrix $\bJ$ from observations.  The likelihood function for the coupling matrix $\bJ'$ given 
the observations is $\mcL(\bJ'|\{s_k\}) = P(\{s_k\}|\bJ')$.
We estimate the network couplings as 
$\widehat{\bJ} = \text{argmax}_{\bJ'} \mcL(\bJ'|\{s_k\})$, the parameter values $\widehat{\bJ}$ that maximize the likelihood $\mcL$ given a set of network state observations $\{s_k\}$.

\section*{\label{sec:infPhys}Information and physics in model inference}
When is accurate inference possible in a spin network? We seek to convert observations drawn from the true equilibrium distribution $P(s_k|\bJ)$ into information about the couplings (spin interactions) $\bJ$ in the underlying network. How much information does $P(s_k|\bJ)$ carry about $\bJ$?  

Fundamentally, inference is impacted by the environment of a network.  For example, $P(s_k|\bJ)$ is strongly dependent on temperature. As $T \rightarrow 0$, $P(s_k|\bJ)$ becomes dominated by ground states that minimize $E(s_k|\bJ)$, and many coupling matrices $\bJ$ have the same ground states. Similarly, at high temperature, all networks achieve the same state distribution $P(s_k|\bJ) \sim 1/2^m$: samples are uniformly distributed across all $2^m$ possible network states, independent of $E(s_k|\bJ)$, and so all networks share the same $P(s_k|\bJ)$. 
These simple arguments suggest an optimal regime for inference at an intermediate temperature, when training data is not just confined to ground states, yet remains distributed non-uniformly to provide information about the relative energy of different states and hence the underlying couplings $\bJ$.

\subsection*{The information content of observations}
A good regime for learning is characterized by a data distribution that changes dramatically for a small change in underlying system parameters. A standard measure for distinguishability---here between the state distribution $P(s|\bJ)$ generated by the network with couplings $\bJ$ and the distribution $P(s|\bJ')$ predicted by the model with inferred couplings $\bJ'$~\cite{watanabe2009algebraic}---is the relative entropy (Kullback-Leibler divergence)~\cite{cover06},
\begin{align}
D[P(s|\bJ) || P(s|\bJ') ] \equiv \sum_k P(s_k|\bJ) \log \tfrac{P(s_k|\bJ)}{P(s_k|\bJ')} \ .
\end{align}
$D[P(s|\bJ) || P(s|\bJ') ] = 0$ if and only if the two distributions are identical. 
When the two distributions are similar ($\bJ' \approx \bJ$), the leading-order contribution to the relative entropy is
\begin{align}
D[P(s|\bJ) || P(s|\bJ') ]
\approx \tfrac{1}{2} \bJ^T \bmcI \bJ \ ,
\end{align}
in terms of the Fisher information matrix~\cite{cover06}, 
\begin{align}
\mcI_{ij,\ell m}(\bJ) \equiv \left\langle \frac{\partial^2 \log \mcL(\bJ|s_k)}{\partial J_{i,j} \partial J_{\ell,m}} \right\rangle_{P(s_k|\bJ)} \ ,
\end{align}
where the $ij,\ell m$-entry gives the mixed partial derivative of the log-likelihood with respect to the coefficient coupling spins $i$ and $j$, and the coefficient coupling spins $\ell$ and $m$.  
The Fisher information measures the curvature of the relative entropy near the true parameters, and it thus provides a quantitative measure of the expected information content of an observation drawn from the equilibrium distribution $P(s_k|\bJ)$. 

For spin networks, 
$\mcI$ has
a simple analytic form: 
\begin{align}
\mcI_{ij,\ell m}(\bJ) 
&= \f{\l\langle \l(\frac{\partial E}{\partial J_{i,j}}\r)\l(\frac{\partial E}{\partial J_{\ell,m}}\r) \r\rangle - \l\langle \frac{\partial E}{\partial J_{i,j}} \r\rangle\l\langle \frac{\partial E}{\partial J_{\ell,m}} \r\rangle}{T^2} \\
&= \f{\l\langle\sigma_i \sigma_j \sigma_{\ell} \sigma_m \r\rangle - \l\langle\sigma_i \sigma_j\r\rangle \l\langle\sigma_{\ell} \sigma_m \r\rangle}{T^2} \\
\mcI_{ij,ij}(\bJ) 
&= \f{\text{Var}[\sigma_i \sigma_j]}{T^2} \ .
\label{eq:Icorr}
\end{align}
The diagonal entry is the ratio of $\text{Var}[\sigma_i \sigma_j]$, the fluctuations in spin-spin alignment within the network, to the squared temperature.  
This reveals a trade-off: greater 
network fluctuations populate excited states to permit the inference of their energies and hence spin couplings, yet also reduce the difference in equilibrium probabilities of states with different energies. The optimal temperature for learning balances these two effects.  


The spin-spin alignment variance is directly related to the spin-spin correlation function $\langle \sigma_i \sigma_j \rangle$ via
$\text{Var}[\sigma_i \sigma_j] = 1 - \langle \sigma_i \sigma_j \rangle^2$.
$\langle \sigma_i \sigma_j \rangle$ has been a central object in the study of spin networks~\cite{MezardMontanari}, providing an operational measure of spin-network ordering~\cite{Schneidman:2006he,MezardMontanari}. 
Moreover, $\langle \sigma_i \sigma_j \rangle$ can be computed directly from observations, permitting estimation of the Fisher information. 

\begin{figure}[t]
\centering
\includegraphics[width=0.5\textwidth]{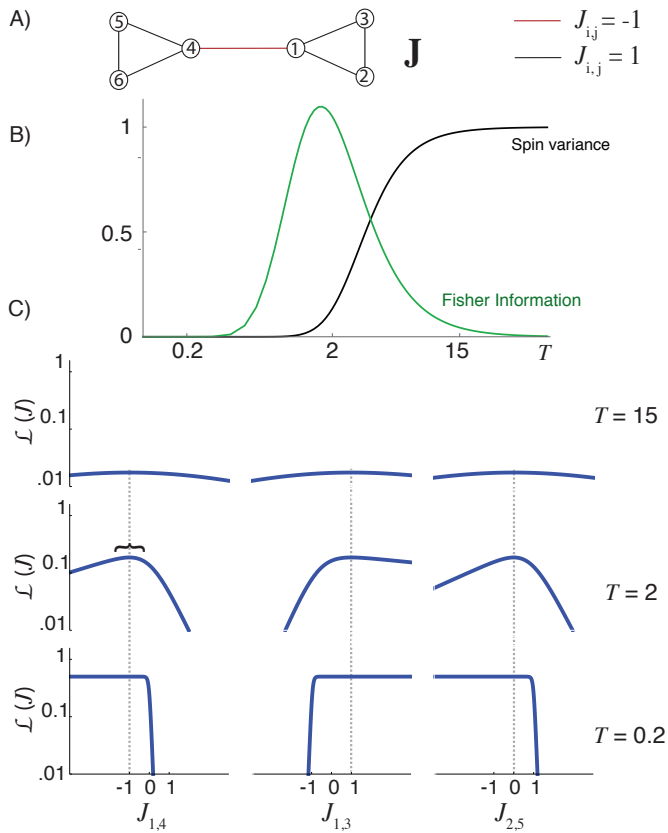}
\caption{Fluctuations and information. 
(A) True network architecture, with black links indicating positive (+1) couplings and red links negative (-1) couplings. 
(B) Minimum diagonal entry $\min_{i,j}\,\mcI_{ij,ij}$ of Fisher information (green curve) and spin-spin variance $\langle (\sigma_i \sigma_j)^2 \rangle - \langle \sigma_i \sigma_j \rangle^2$ (black curve), averaged over spin pairs in true network $\bJ$. $\min_{i,j}\mcI_{ij,ij}$ occurs for the negative coupling, $i=1$ and $j=4$. 
(C) Likelihood as function of coupling coefficient $J_{i,j}$, for high ($T=15$, top row), medium ($T=2$, middle), and low temperatures ($T=0.2$, bottom). Three coupling coefficients are shown, that in the true network are negative ($J_{1,4}$, left), positive ($J_{1,3}$, middle) or zero ($J_{2,5}$, right). Curly bracket (middle left) highlights at intermediate temperature the large curvature of the likelihood, and hence large Fisher information, at the true parameter value.
}
\label{fig:fluctInf}
\end{figure}

Figure~\ref{fig:fluctInf} illustrates the relationship between spin correlations and information for a simple model spin network containing two groups of spins, with positive (weight +1) intra-group couplings and a single negative (weight -1) inter-group coupling (Fig.~\ref{fig:fluctInf}a).
As $T \to 0$, the spin-spin variance approaches $0$
and hence Fisher information is low (Fig.~\ref{fig:fluctInf}b): the network is frozen in a 
small
set of ground states that carry insufficient information for inference.  
As temperature increases, network fluctuations increase and the Fisher information achieves a maximum value at an intermediate temperature. 
As $T \to \infty$, inference again becomes degenerate, as temperature-driven fluctuations dominate.

Since Fisher information 
quantifies
the curvature of the likelihood $\mcL$ near the true network, the three regimes of inference are directly reflected in the geometry of $\mcL$ near the true value of $\bJ$ (Fig.~\ref{fig:fluctInf}c).  At low and high $T$, low Fisher information manifests as a flat $\mcL$ 
near the true coupling coefficient $J_{i,j}$:
inference is degenerate and many parameter values have equivalent likelihood given data. For example, $J_{2,5}$ can equal $-1,0,1$ with equivalent values of $\mcL$. By contrast, at intermediate $T$ the peak in Fisher information 
is reflected in the strong curvature of 
$\mcL$ 
at its
local maximum. 

\subsection*{Information and sampling complexity}
$\mcI_{ij,ij}$ provides a direct measure of inference efficiency. Specifically, the Cramer-Rao bound relates the Fisher information to a lower bound on the 
average L2
inference error, defined as the mean-squared error between $\bJ$ and an unbiased estimator $\widehat{\bJ}$~\cite{cover06}. 

The Cramer-Rao bound is generally stated for the inverse of the Fisher information matrix, however a convenient loose bound is~\cite{bobrovsky1987some}
\begin{equation}
\l\langle \l(\widehat{J}_{i,j}(\{s_k\}) - J_{i,j}\r)^2\r\rangle_{P(\{s_k\}|\bJ)} \ge \frac{1}{N \mcI_{ij,ij}(\bJ)} \ ,
\end{equation}
for $N$ observations drawn from $P(s_k|\bJ)$. 
In this way, $\mcI$ quantifies the information that observations provide about the network couplings $\bJ$. 

For inference error (averaged over unique spin pairs $\{i,j\}$)
$\epsilon \equiv 
\frac{2}{m^2 - m}
\sum_{\{i,j\}}\l\langle\l(\widehat{J}_{i,j}(\{s_k\}) - J_{i,j}\r)^2 \r\rangle_{P(\{s_k\}|\bJ)}$,
the harmonic mean $\mcI_{\rm hm}(\bJ) \equiv \l[\frac{m^2-m}{2}\sum_{\{i,j\}} \mcI_{ij,ij}^{-1}(\bJ)\r]^{-1}$
of diagonal Fisher information matrix entries provides a direct measure of the sampling complexity, defined in machine learning as the number of samples $N_{\epsilon_0}$ required to achieve a desired bound $\epsilon_0$ on the average error:
\begin{align}
N_{\epsilon_0} &\ge 
\f{1}{\epsilon_0 \mcI_{\rm hm}(\bJ)} \ .
\end{align}
For one significantly smaller diagonal entry, the harmonic mean is approximately the minimum,
$\mcI_{\rm hm}(\bJ) \approx \min_{i,j}\mcI_{ij,ij}(\bJ)$.

\section*{\label{sec:ModelArchitectures}Inference in model spin-network architectures}
\subsection*{\label{sec:SpinChains}Optimal inference for 1D spin chains}

Analytic forms for $\langle \sigma_i \sigma_j \rangle$ (and hence the Fisher information) cannot in general be directly computed, however some simple networks do permit analytic expressions. An interesting special case is the 1-dimensional spin chain---a network with $J_{i,i+1} = 1$, and $J_{i,j}= 0$ otherwise---which has simple correlations 
$\langle \sigma_i \sigma_j \rangle_{P(s_k | \bJ)} = \l( \tanh \f{1}{T} \r)^{|i-j|}$
and hence Fisher information~\cite{MezardMontanari}:
\begin{align}
\mcI_{ij,ij}(\bJ) &= \f{1 - \l( \tanh \f{1}{T} \r)^{2 |i-j|}}{T^2} \ .
\label{eq:1DIsing_FI}
\end{align}

\begin{figure}
\centering
\includegraphics[width=0.5\textwidth]{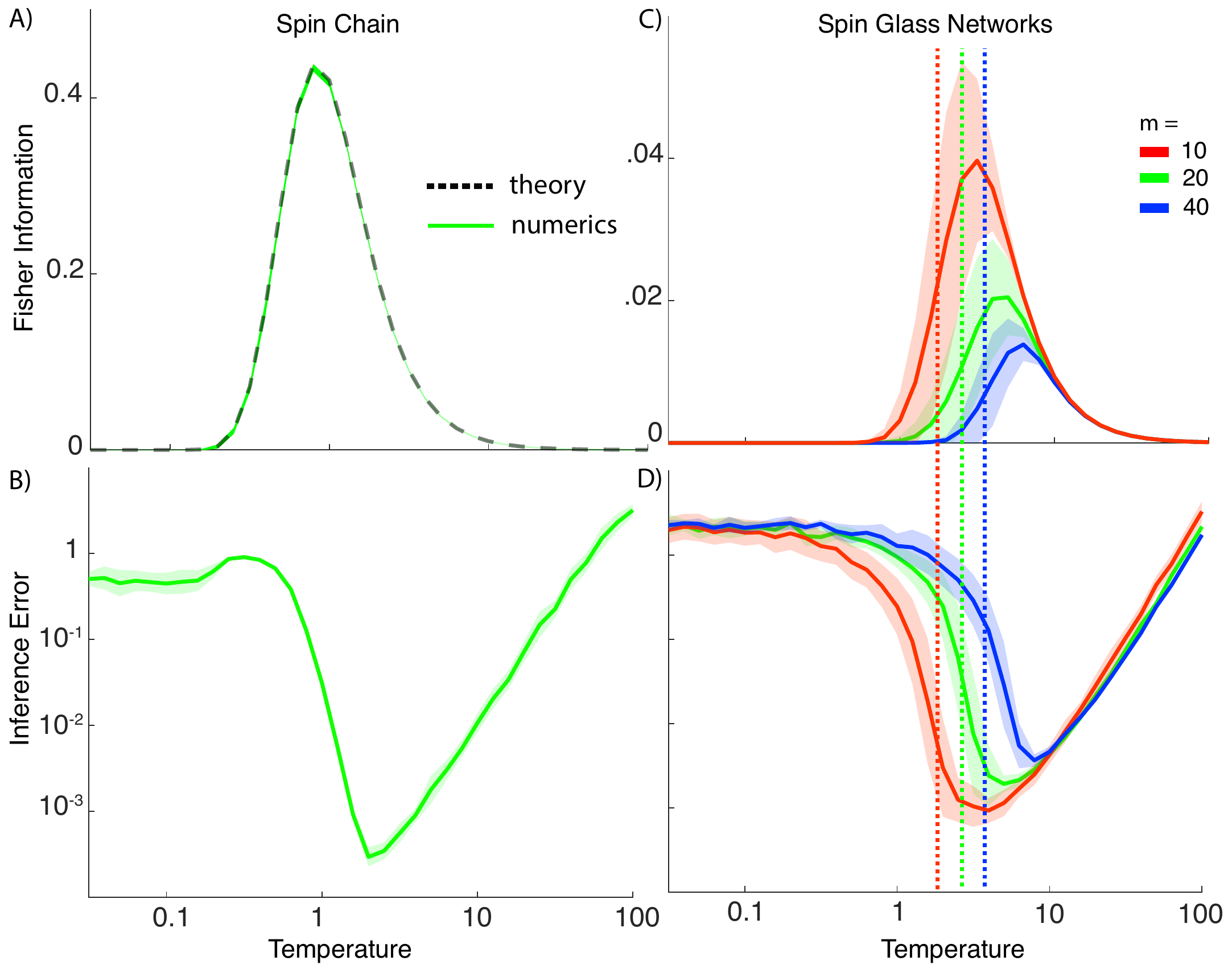}
\caption{Inference in example network architectures. Minimum diagonal entry $\min_{i,j}\mcI_{ij,ij}$ of the Fisher information matrix (top row) and L2 inference error (bottom row) versus temperature for uniform 1D chain 
(left column) and spin-glass networks 
(right column). 
(A) Theoretical expression for Fisher information~[\eqref{eq:1DIsing_FI}] (black dashed curve) closely matches numerical estimation
(green solid curve). 
(B) Inference error from likelihood maximization on $m=20$ chain, as a function of temperature. Green curve and shaded region show mean $\pm$ standard deviation for $50$ 
different samplings. Each replicate is calculated using different random samples drawn from $P(s_k | \bJ)$.
(C) 
Fisher information
for spin-glass network size $m=$10 (red), 20 (green), or 40 (blue). 
Dark curves and shaded regions show means $\pm$ standard deviations for 50 different networks, and vertical dotted lines show corresponding $\Tc$. 
(D) Inference error for likelihood maximization on corresponding networks.
} 
\label{fig:fisherError}
\end{figure}

Figure~\ref{fig:fisherError}a shows the minimum diagonal entry of the Fisher information matrix increasing as a function of temperature, obtaining a maximum at 
an intermediate temperature, 
and decaying asymptotically to zero as $T \rightarrow \infty$. Similarly, maximum-likelihood inference error is minimized at an intermediate temperature (Fig.~\ref{fig:fisherError}b). 
Results are theoretically identical and numerically indistinguishable for spin chains of different lengths (data not shown).

\eqref{eq:1DIsing_FI} shows that the average information content of observations increases as a function of distance $|i-j|$ between spins, because $\langle \sigma_i \sigma_j \rangle_{P(s_k | \bJ)}$ decays exponentially with distance. 
Thus, long-range links are `easier' to learn because correlations between distant spins are weaker. Nearby links are harder to learn because correlations are strong and less information accrues per observation. In the 1D spin chain, the challenge of inference is dominated by the effort of learning the coupling $J_{i,i+1}$ of the most highly correlated nearest-neighbor spins. 

\subsection*{\label{sec:SpinGlass}Optimal inference in spin-glass networks}
For specific random ensembles of $\bJ$, we can numerically evaluate the Fisher information. An important class of spin networks are Gaussian random spin glasses, where couplings are Gaussian-distributed according to $P(J_{i,j}) \sim \mathcal{N}(0;K^2)$, a normal distribution with mean $0$ and variance $K^2$. 
These networks have both positive and negative couplings and rich physical behaviors, so have been a model system for studying complex interacting systems. 
We fix $K=1$ to set the energy scale in our system, and hence determine the scale of $T$ needed to achieve a given error.

Figure~\ref{fig:fisherError}c shows the minimum diagonal entry of the Fisher information matrix, averaged over 50 randomly sampled spin-glass networks each of size $m=10, 20$, and $40$.
Consistent with our earlier asymptotic arguments, 
Fisher information displays 
a local maximum 
as a function of $T$.
Moreover, the maximum in $\mcI_{ij,ij}(\bJ,T)$ approximately coincides with the minimum in inference error (Fig.~\ref{fig:fisherError}c).

This optimal inference temperature $\Topt$ appears to track the critical temperature $\Tc = \frac{1}{2} \sqrt{\frac{K^2 m}{\ln 2}}$ of a well-characterized phase transition in the Gaussian spin glass in the absence of a magnetic field $h$~\cite{Derrida:1980eda,MezardMontanari} (vertical dotted lines in Fig.~\ref{fig:fisherError}c,d).
Below this critical temperature, the network's state distribution is concentrated on a small number of low-energy states, and for finite samples explores only an exponentially small fraction of possible configurations. Near the phase transition, the state distribution $P(s_k)$ transitions abruptly from being `frozen' (sharply peaked on a small number of ground states) to being dispersed, and there is a correspondingly abrupt change in the spin-spin correlations, the Fisher information, and the efficiency of inference.

\section*{\label{sec:ActiveLearning}Optimizing inference with active learning}
In general, the optimal temperature for inference, like $\Tc$ for Gaussian networks, depends on parameters including the network topology and the variance of network couplings. Such network-specific parameters, and thus the optimal temperature for inference, are generally unknown \textit{a priori} by an observer.
This motivates the development of active learning paradigms that can determine $\Topt$ through interaction with the system.

We design a learning protocol that operates in two stages. In the first stage, we directly manipulate temperature to optimize inference efficiency.
We estimate the optimal temperature $\Topt$ by empirically maximizing the minimum estimated Fisher information across all coupling coefficients,
$\widehat{T}_{\rm opt} \equiv \text{argmax}_{T} \left[ \min_{i,j} \widehat{\mcI}_{ij,ij}(\bJ) \right]$.
Importantly, the spin-network Fisher information can be estimated on-line via the empirical spin-spin correlation function 
$\f{1}{N} \sum_{\{s_k\}} \sigma_i^{(k)} \sigma_j^{(k)}$
sampled during an initial temperature `sweep.'  
In the second stage, we use maximum-likelihood estimation (see Materials and Methods) to infer $\bJ$ from observations collected at $\widehat{T}_{\rm opt}$, the inferred optimal temperature for inference. 

\begin{figure}
\centering
\includegraphics[width=0.5\textwidth]{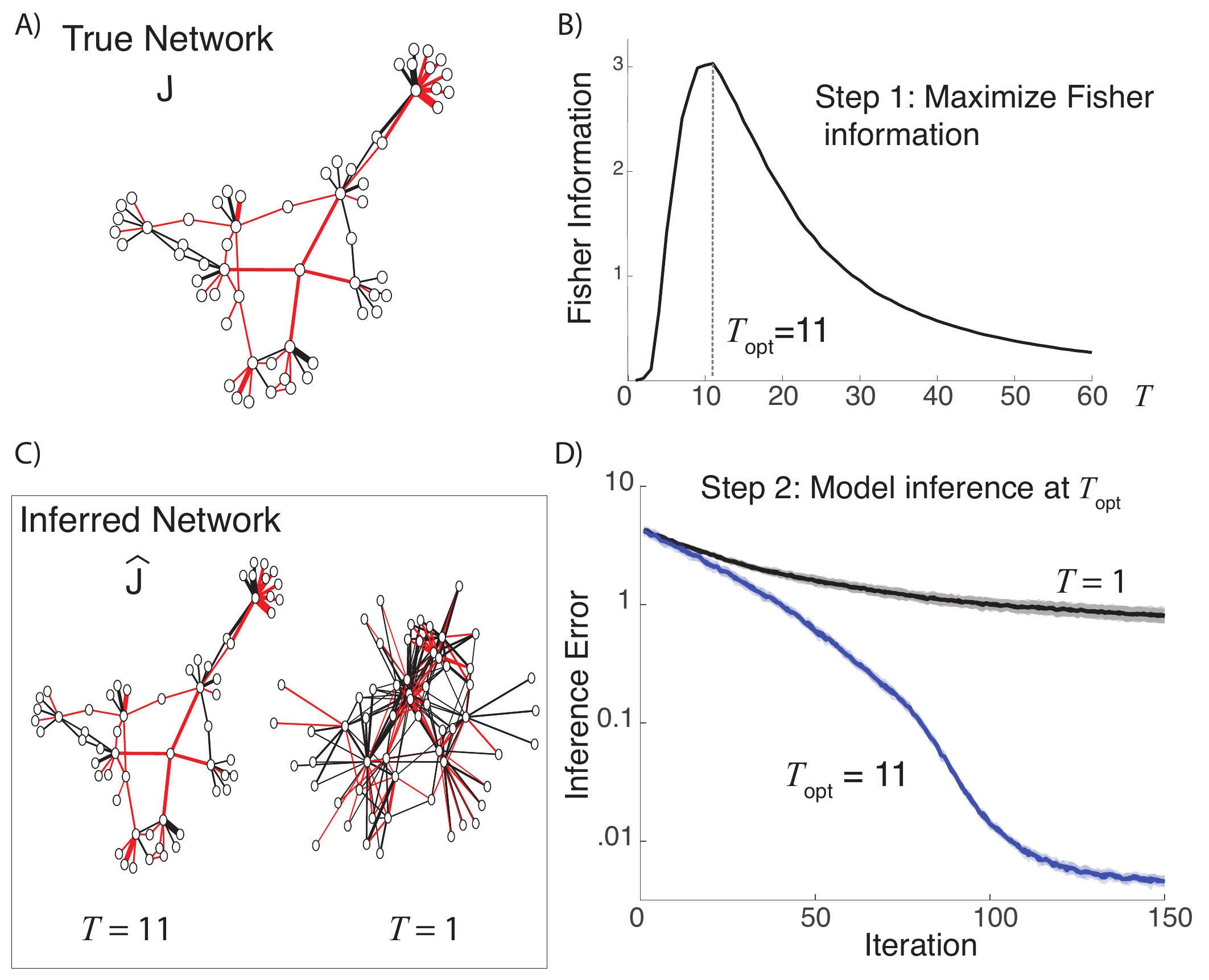}
\caption{
Active inference. 
(A) Spin network modeled on mouse cortex reconstruction from \cite{Bock:2011fk}. Edges are red for $J_{i,j}<0$ and blue for $J_{i,j}>0$.
(B) Minimum diagonal entry $\min_{i,j}\mcI_{ij,ij}(\bJ,T)$ of Fisher information matrix, as function of temperature, 
is maximized at $T = 11$. 
(C) Network reconstructed through likelihood maximization 
from equilibrium samples collected at $T=11$ (left) 
or 
at $T=1$ (right).
Links omitted for $|J_{i,j}|<0.1$.
(D) L2 inference error as function of iteration for $T=11$ (blue) and $T=1$ (black). 
Dark curve and shaded region show mean $\pm$ standard deviation from 10 runs. 
}
\label{fig:1D Active}
\end{figure}

To test an interesting case, we selected a relatively large network (with $m=60$ nodes) based on a 
reconstruction of a cortical brain circuit~\cite{Bock:2011fk} (Fig.~\ref{fig:1D Active}a). Figure~\ref{fig:1D Active}b shows the initial estimation of Fisher information as a function of temperature,
which found a global maximum near $T=11$.
Maximum-likelihood inference from equilibrium samples collected at $T=11$ correctly identified the topology of the network (Fig.~\ref{fig:1D Active}c) and produced an L2 error 
$\epsilon <.01$ per coupling (Fig.~\ref{fig:1D Active}d). By contrast, inference at $T = 1$ produced a network with many erroneous linkages and uniformly high error. 
In this way, temperature optimization allows efficient learning of large networks. 

\section*{\label{sec:Discussion}Discussion}
In this paper, we studied fundamental limits imposed by the physical behavior of a spin network on the learning of predictive models. We showed that the physics of a spin network strongly affects the efficiency of inference. Specifically, fluctuations play a particularly important role in determining when inference is possible. 

Intuitively, inference is degenerate when many different networks produce similar distributions over states.  This occurs generically both at low temperature, when a network explores a small number of configurations and is focused on ground states, and at high temperature when a network samples all states equally. Thus, below an optimal temperature, learning efficiency is enhanced by increasing the temperature, thereby increasing the strength of the thermal noise.
Greater fluctuations drive the network to explore a broader range of states, thereby enabling the inference of sufficiently constrained models.  Above the optimal temperature, additional thermal noise becomes detrimental to learning: when random driving is too high, couplings are hidden by the dominance of noise.

Our results reveal a physical dimension of inference and machine learning. In machine learning, standard paradigms---including PAC learning~\cite{Valiant:1984kt} and Empirical Risk Minimization~\cite{Poggio:gt,vapnik}---relate the complexity of the learning task to the statistical properties of training data. In PAC learning, for example, the probability of obtaining training samples from a class of interest determines the number of samples required to meet a desired error bound. In physical systems, physical variables (like network temperature) impact the statistical properties of the training data, and thus, physical variables modulate the probability of obtaining training examples that allow accurate learning and model generalization. This observation provides a fundamental link between physics and machine learning.

Here, we argue that for a spin network there are optimal environmental conditions for learning. 
If this result holds more generally, a natural network of biological interest---such as a neural network or gene regulatory network---might benefit from experimental strategies that inject noise into the system. 
Further, there might be `optimal' noise magnitudes when inference becomes most tractable.

Experimentally, the success of our simple active-learning methodology 
motivates feedback protocols for interacting with a system to bring it near conditions of optimal inference.  The framework we describe here would require closed-loop monitoring of spins and a `noise source.' However, we anticipate that broader strategies for optimizing inference might include 
additional classes of 
perturbations to a system; for example, in spin networks one could manipulate magnetic field as well as temperature. 

Originally developed in statistical mechanics to model phase transitions in simple magnetic materials, spin-network models have provided useful descriptions of non-equilibrium systems such as bird flocks and neural circuits~\cite{Schneidman:2006he,Hinton:2006kc,Hopfield:1982vm,Katz:2011fb,Cocco:2009hw}. 
The broad utility of these models suggests that our results may provide an avenue to develop `optimal inference protocols' for the construction of predictive models in a wide range of domains beyond the formal scope of equilibrium physics, including problems in chemistry, biology, and ecology.


\begin{acknowledgments}
This work was supported by grants from the NIH (NIH DP5 OD01219), the CZI, Amgen, and Rosen Center at Caltech (MT); a Natural Sciences and Engineering Research Council of Canada (NSERC) Discovery Grant, the Canada Research Chairs program, and the Faculty of Science, Simon Fraser University through the President's Research Start-up Grant (DAS); and WestGrid (www.westgrid.ca) and Compute Canada Calcul Canada (www.computecanada.ca). The authors thank Andrew Stuart, David van Valen, Joel Tropp, Rob Phillips, John Doyle, Carl Pabo, Malcolm Kennett, Emma Lathouwers, and Erik Winfree for useful discussions and feedback on the manuscript.
\end{acknowledgments}

\section*{\label{sec:Methods}Method Details}
In the numerical examples shown in Figs.~\ref{fig:fisherError} and \ref{fig:1D Active}, we perform maximum-likelihood estimation of spin-network architectures from data. Briefly, we calculate the estimator $\widehat{\bJ}$ of the coupling matrix by numerically maximizing $\mcL(\bJ'|\{s_k\})$ via gradient ascent, given state observations $\{s_k\}$ drawn from $P(s_k|\bJ)$, the equilibrium state distribution of the `true' network with couplings $\bJ$. 

Explicitly, each iteration of the learning algorithm applies to the current estimate $\bJ(n)$ a simple update rule, using the average over $N$ observations of the gradient of the log-likelihood $\log \mcL$: 
\begin{align}
\label{eq:ascent}
J'_{i,j}&(n+1) = J'_{i,j}(n) +  \eta \ 
\frac{\partial}{\partial J'_{i,j}} \l[ \f{1}{N} \sum_{\{s_k\}} \log \mcL(J'_{i,j}|s_k) \r] \\
&= J'_{i,j}(n) + \f{\eta}{T} \left(\f{1}{N} \sum_{\{s_k\}} \sigma_i^{(k)} \sigma_j^{(k)} - \langle \sigma_i \sigma_j \rangle_{P(s_k|\bJ')}\right) \ . 
\end{align}
Here, $\langle \sigma_i \sigma_j \rangle_{P(s_k|\bJ')}$ is the spin-spin correlation for equilibrium fluctuations under the estimated coupling $\bJ'$, and $\tfrac{1}{N} \sum_{\{s_k\}} \sigma_i^{(k)} \sigma_j^{(k)}$ is the empirical spin-spin correlation under the true coupling $\bJ$. 
$\eta$ is the \emph{learning rate} parameter that sets the update size. In our optimization, we use $\eta = \frac{2}{1+\f{0.1}{z}}$ where $z$ is the iteration number, so that our learning rate decays during the gradient ascent~\cite{hinton2003stochastic}.

Observations are sampled from the equilibrium distribution $P(s_k|\bJ)$ with Metropolis sampling. In general the $N$ required to estimate spin-spin correlations is a function of network size, and the accuracy of correlation function estimation scales with increasing $N$. We selected $N=50,000$ in our model systems.  

\bibliography{references}



\end{document}